\documentclass[aps,prl,twocolumn,amsmath,amssymb,showpacs,showkeys]{revtex4}
\usepackage{graphicx}
\usepackage{dcolumn}
\usepackage{bm}

\begin{document}
\title{ A variational principle for the {P}areto power law }

\author{Anirban Chakraborti}
  \email{anirban.chakraborti [at] ecp.fr}
  \affiliation{
  Laboratoire de Math\'{e}matiques Appliqu\'{e}es aux Syst\'{e}mes, 
  Ecole Centrale Paris, 92290 Ch\^{a}tenay-Malabry, France}

\author{Marco Patriarca}
  \email{marco.patriarca [at] kbfi.ee}
  \affiliation{National Institute of Chemical Physics and Biophysics, 
  R\"avala 10, 15042 Tallinn, Estonia\\
  IFISC, Instituto de F\'isica Interdisciplinar y Sistemas Complejos (CSIC-UIB), 
  E-07122 Palma de Mallorca, Spain
}

\date{\today}

\begin{abstract}
A mechanism is proposed for the appearance of power law distributions 
in various complex systems.
It is shown that in a conservative mechanical system composed of subsystems with 
different numbers of degrees of freedom a robust power-law tail can appear in 
the equilibrium distribution of energy as a result of certain superpositions 
of the canonical equilibrium energy densities of the subsystems.
The derivation only uses a variational principle based on the Boltzmann 
entropy, without assumptions outside the framework of canonical equilibrium 
statistical mechanics.
Two examples are discussed, free diffusion on a complex network and 
a kinetic model of wealth exchange.
The mechanism is illustrated in the general case through an exactly 
solvable mechanical model of a dimensionally heterogeneous system.
\end{abstract}





\maketitle

\emph{Introduction.}
The study of plausible mechanisms responsible for the appearance of power law distributions,
observed so frequently in Nature, represents an active research field.
Power laws were first met in fields far from traditional physics,
e.g. Pareto's power law of income distribution in economics and 
the Zipf law for the occurrence rate of words in linguistics~\cite{Newman2005a},
suggesting the existence of general characteristics for the underlying dynamics.
Power law distributions are also found in systems out of equilibrium,
e.g. with avalanche dynamics~\cite{Turcotte1999a} or in
multiplicative stochastic processes~\cite{Sornette1998a}.
Their peculiar character, strikingly different respect to the exponential tails
of canonical distributions predicted by equilibrium statistical mechanics,
challenges our intuition.

A first-principles derivation of equilibrium distributions with a power law 
tail is possible at the price of additional hypotheses on the foundations of
canonical equilibrium theory, e.g. in non-extensive generalizations 
of the Boltzmann entropy~\cite{Tsallis1988},
or modifying the basic form of the Gibbs distribution~\cite{Treumann2008a}. 
Power laws are also obtained within the framework of 
superstatistics~\cite{Beck2006a}, 
if some parameter, typically the temperature,  is considered as a random 
variable.

The present paper introduces a novel mechanism based on the heterogeneity of 
the system dimensionality, characterizing systems composed of subsystems 
with different numbers $N$ of degrees of freedom distributed according 
to a dimension density $P(N)$.
The derivation uses a variational approach based on the Boltzmann entropy, with 
no additional assumptions outside canonical statistical mechanics.
A robust power law tail appears due to the superposition of the canonical 
equilibrium distributions of subsystems with high dimensions $N \gg 1$.

The phenomenon investigated here resembles how in the gene 
regulation network model of Balcan and Erzan~\cite{Mungan2005a}
a scale-free network results from the superposition of Poissonian 
out-degree distributions.
Also, a high phase space dimensionality is known to play a role 
in the anomalous dynamics of some spin-glass models~\cite{Kurchan1996a}.

The mechanism proposed here for the generation of power law tails
is best illustrated for systems which can be clearly defined
as dimensionally heterogeneous, on the basis of 
a mechanical analogy or a geometrical interpretation.
For this reason, two explicit examples of such systems are presented.
Then the mechanism is formalized in terms of a simple yet general 
dimensionally heterogeneous mechanical model. 

\emph{Diffusion on networks.}
As a first example we compare a free diffusion process on a homogeneous network 
and that on a scale-free network.
We first consider a homogeneous lattice of $M$ sites, in which each site $i$ 
($i=1,\dots,M$) is connected to the same number $k_i = k$ of first neighbors.
Such a lattice is an example of a dimensionally homogeneous network providing a 
discrete representation of a $N$-dimensional space.
In the case of the square lattice structure, the dimension $N$ is related to 
the degree $k$ as $N = k/2$.
An unbiased uniform diffusion process of $X$ walkers hopping between 
the $M$ sites of the lattice relaxes toward a uniform load distribution
$f(x) = \mathrm{const}$, with the same average load at each node $i$ given by 
$x_i = X/M$. 
On the other hand, a heterogeneous network with degree distribution $g(k)$ 
cannot be given a straightforward geometrical interpretation
and in general represents a space with a highly complex topology. 
In particular, no unique dimension can be assigned, so that it can be regarded 
as a dimensionally heterogeneous space.
One can estimate a local dimensionality from the connectivity, in analogy with 
the homogeneous square lattice, by introducing for each node $i$ the local 
dimension $N_i = k_i/2$~\cite{note1}.
At equilibrium, free diffusion of $X$ walkers on such a network is known to 
relax to an average load $x_i$ proportional to the degree~\cite{Noh2004a}, 
$x_i = \bar{x} k_i$, where the average flux per link and direction $\bar{x}$
is fixed by normalization, 
$\bar{x} = X/K$, with $X = \sum_{i} x_i$ and $K = \sum_{j} k_j$.
It follows from probability conservation that the load distribution 
at equilibrium is directly determined by the degree distribution,
\begin{eqnarray}  
  \label{PQ}
  f(x) = g(k) dk/dx = g(x/\bar{x}) / \bar{x} \, .
\end{eqnarray}
In the important case of a scale-free network with 
$g(k \!\gg\! 1) \sim 1/k^p$, one has a power law tail in the load distribution,
$f(x) \sim 1/x^p$, with the same exponent $p$.
A close relation between degree distribution and equilibrium density,
analogous to Eq.~(\ref{PQ}) valid for the case of free diffusion,
can be expected for any quantity $x$ diffusing through a network. 
For instance, in the case of the Zipf law, such a relations is known to hold
approximately. 
In this case written language is regarded as a random walk across the complex 
network with nodes given by words and links between words which are neighbors 
in a text~\cite{Masucci2009a}.

\emph{Kinetic exchange models.}
A second example of dimensionally heterogeneous system is suggested by 
kinetic exchange models of wealth (KWEM)~\cite{Chatterjee2007b}.
In KWEMs, $M$ agents exchange wealth through pair-wise interactions.
At each time step two agents $i,j$ are extracted randomly and their wealths 
$x_i,x_j$ updated depending on the saving parameters $\lambda_i,\lambda_j$, 
representing the minimum fractions of wealth saved~\cite{Chakraborti2000a},
\begin{eqnarray}  \label{sp1b}
  x_i' 
  &=& 
  \lambda_i x_i + \epsilon [(1-\lambda_i) x_i + (1-\lambda_j) x_j]  \, ,
  \nonumber \\
  x_j' 
  &=& 
  \lambda_j x_j - (1 - \epsilon) [(1-\lambda_i) x_i + (1-\lambda_j) x_j ] \, .
\end{eqnarray}
Here $x_i',x_j'$ are the wealths after a trade and $\epsilon$ a uniform random 
number in $(0,1)$.
Notice that $x$ is conserved during each transactions, $x_i+x_j=x_i'+x_j'$.
This dynamics closely recalls energy exchange in molecular collisions,
an analogy noticed by Mandelbrot~\cite{Mandelbrot1960a} which becomes 
closer introducing an effective dimension associated to the system:
in the homogeneous version of KWEMs, i.e. all $\lambda_k \equiv \lambda$,
the equilibrium wealth distribution $f(x)$ is well fitted by the energy 
distribution of a perfect gas in a number $N$ of dimensions defined by 
$N(\lambda)/2= 1 + 3 \lambda/(1 - \lambda)$, i.e. a 
$\Gamma$-distribution $\gamma_{N/2}(x)$ of order $N(\lambda)/2$~\cite{Patriarca2004a}.
In fact, inverting $N(\lambda)$, one obtains an average fraction of wealth 
exchanged during one trade $1 \!-\! \lambda \!\propto\! 1 / N$ ($N \!\gg\! 1$),
similarly to the energy exchanges during molecular collisions of 
an $N$-dimensional gas~\cite{Chakraborti2008a}.
Then a heterogeneous system composed of agents with different $\lambda_i$ is 
analogous to a dimensionally heterogeneous system.
The relevance of heterogeneous versions of KWEMs with $\lambda_i$ distributed 
in the interval $(0,1)$  is in the fact that they relax toward realistic wealth 
distributions $f(x)$ with a Pareto tail, as shown numerically and 
analytically~\cite{Chatterjee2007b}.
In the case of a uniform distribution for the saving parameters,
$\phi(\lambda) = 1$ if $\lambda \in (0,1)$ and $\phi(\lambda) = 0$ otherwise, 
setting $n=N/2$, the dimension density has a power law $\sim 1 /n^2$, 
$P(n) = \phi(\lambda) d\lambda/dn = 3/(n + 2)^2$ ($n \!\ge\! 1)$.

\emph{A dimensionally heterogeneous mechanical model.}
The power law tails of the equilibrium distributions in the examples 
presented above are particular cases of a more general mechanism which takes
place in dimensionally heterogeneous systems.
We illustrate it through a simplified yet general exactly solvable mechanical 
model, which can be interpreted as an assembly of polymers with different 
numbers of harmonic degrees of freedom, representing e.g. small displacements 
of its normal modes respect to the equilibrium configuration.
For simplicity no interaction term between polymers is included in the total 
energy function, analogously to the statistical mechanical treatment of the 
molecules of a perfect gas.
The absence of interactions between units implies that each subsystem undergoes 
independent statistical fluctuations, while a more rigorous treatment of e.g. 
diffusion on a complex network or wealth exchange in the KWEMs considered above
should take into account the correlations between nodes or agents.
However, the mechanical model studied here below preserves the basic features
of the heterogeneous dimensionality and statistical character due to internal 
random fluctuations.

First, we recall the results for a dimensionally homogeneous systems, e.g. 
an assembly of identical polymers $N$ degrees of freedom~\cite{Patriarca2004a}.
For a quadratic energy function $x(\mathbf{q}) = (q_1^2+\dots+q_N^2)/2$,
where the $q_i$'s are the $N$ variables of the polymer, the equilibrium density 
is obtained varying respect to $f_n(x)$ the functional $S_n[f_n]$, 
obtained from the Boltzmann entropy~\cite{Patriarca2004a},
\begin{eqnarray} \label{func6b}
  \!\!\!\!\!S_n[f_n] =
  \int_{0}^{\,+\infty} \!\!\! dx \, f_n(x)
  \left\{ \ln\left[ \frac{f(x)}{\sigma_{2n} \, x^{n-1}} \right]
  + \mu + \beta x \right\}
  \, .
\end{eqnarray}
Here we have introduced the dimension variable
\begin{equation} \label{n}
  n = N / 2 \, , 
\end{equation}
$\sigma_{N}$ is the hypersurface of a unitary $N$-dimensional sphere,
and $\mu, \beta$ are Lagrange multipliers determined by the constraints 
on the conservation of the total number of subsystems and energy, respectively.
The result is a $\Gamma$-distribution of order $n$,
\begin{eqnarray}\label{f-global}
  f_n(x)
  = \beta\,\gamma_{n}(\beta x)
  \equiv \frac{\beta}{\Gamma(n)}(\beta x)^{n-1} \exp(-\beta x) \, ,
\end{eqnarray}
where $\beta^{-1} = 2\langle x \rangle/N = \langle x \rangle/n$, 
according to the equipartition theorem, 
is the temperature~\cite{Patriarca2004b}.

A heterogeneous system is composed of polymers with different numbers of 
degrees of freedom $N_1, N_2, \ldots$ in constant proportions $P_1, P_2, \dots$, 
with $\sum_i P_i = 1$, or, considering a continuous dimension variable $n=N/2$,
according to a distribution $P(n)$, with $\int dn \, P(n) = 1$.
At equilibrium, each subsystem with dimension variable $n$ will have its
probability density $f_n(x)$.
We are interested in the shape of the aggregate equilibrium energy distribution, 
i.e. the relative probability to find a subsystem with energy $x$ 
independently of its dimension $n$.
The equilibrium problem for the heterogeneous system is solved analogously,
from the functional $S[\{f_n\}]$ obtained summing the homogeneous functionals 
with different $n$,
\begin{equation} \label{func7}
  S[\{f_n\}] \! = \! \!
  \int \!\! dn \, P(n) \! \!
  \int_{0}^{+\infty} \! \! \! \! \! \!\! dx \, f_n(x)
  \! \left\{ \! \ln \! \left[ \frac{f_n(x)}{\sigma_{2n} \, x^{n-1}} \right] \!
  \! + \! \mu_n \! + \! \beta x \! \right\} \!  .
\end{equation}
Notice that there is a different Lagrange multiplier $\mu_n$ for each $n$,
since the fractions $P(n)$'s are conserved separately, but
a single $\beta$ related to the total conserved energy.
The equilibrium probability density $f_n(x)$ for the subsystem $n$ is obtained 
by varying $S[\{f_n\}]$ respect to $f_n(x)$ and is again given by
Eq.~(\ref{f-global}), with $\beta$ determined by the total energy,
\begin{eqnarray} \label{x_glob}
  \langle x \rangle
  = \int dn \int_0^{\infty} dx \, f_n(x) \, x
  = \frac{\langle N \rangle}{2\beta} \, ,
\end{eqnarray}
where 
we have introduced the average dimension
\begin{eqnarray} \label{N}
  \langle N \rangle = 2 \langle n \rangle = 2 \int dn \, P(n) \, n \, .
\end{eqnarray}
Equation~(\ref{x_glob}) represents a {\it generalized equipartition theorem}
for dimensionally heterogeneous systems.
To ensure a finite $\langle N \rangle$ (and therefore a finite average energy 
$\langle x \rangle$) the dimension density $P(n)$ has to have a finite cutoff or
decrease faster than $1/n^2$ for $n \!\gg\! 1$.
Using Eq.~(\ref{f-global}), the aggregate distribution is finally
\begin{equation} \label{fn3}
  f(x)
  = \!\! \int \!\! dn \, P(n) f_n(x)
  = \!\! \int \!\! dn \,  
    \frac{P(n)\beta}{\Gamma(n)} (\beta x)^{n-1} 
    \exp(-\beta x) .
\end{equation}
%
\begin{figure}[]
  \centering
  \includegraphics[width=3.0in]{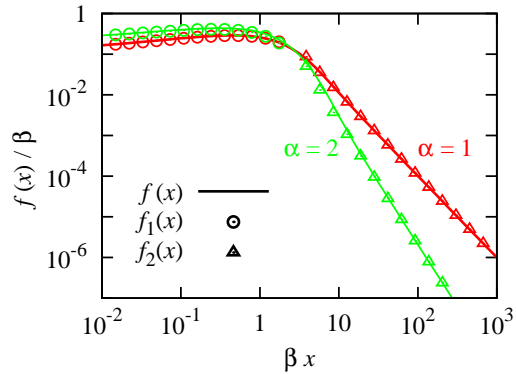}
  \caption{
  \label{fig_1}
  Aggregate distribution $f(x)$, Eq.~(\ref{fn3}),
  with $P(n) = \alpha / n^{1 + \alpha}$ ($n \ge 1$), $P(n) = 0$ otherwise,
  for $\alpha=1$ (red, dark grey), $\alpha=2$ (green, bright grey).
  Continuous lines: numerical integration of Eq.~(\ref{fn3}).
  Triangles: saddle point approximation $f_2(x)$, Eq.~(\ref{f2}).
  Circles: small-$x$ limit, $f_1(x)$, Eq.~(\ref{f1}).
  }
\end{figure}
While the distributions $f_n(x)$ have an exponential tail, 
the function $f(x)$ may exhibit a slower decay and possibly a power law tail, 
if the dimension density $P(n)$ decreases slow enough.
In the example of the power law density $P_\alpha(n) = \alpha / n^{1 + \alpha}$ 
($n \ge 1, \alpha > 0$), $P_\alpha(n) = 0$ otherwise, a power law tail appears 
in $f(x)$, see the continuous curves in Fig.~\ref{fig_1} obtained by 
numerical integration.

In fact, a general result holds, namely an actual equivalence between the 
asymptotic form of the aggregate distribution $f(x)$ and the dimension density 
$P(n)$, whenever $P(n)$ decreases at large $n$ faster 
than $1/n$, expressed by $f(x \!\gg\! \beta^{-1}) \!\!=\!\! \beta P(\beta x)$.
This asymptotic relation can be compared with the equality between the
the average load and the degree distribution, 
$f(x) = g(x/\bar{x}) / \bar{x}$, obtained for the example of free diffusion
on a network with degree distribution $g(k)$.

To demonstrate this relation, we start considering a value $\beta x \gg 1$ 
in Eq.~(\ref{fn3}). 
The main contributions to the integral come from values 
$n \approx \beta x \gg 1$, since $\gamma_{n}(\beta x)$ has its maximum at 
$x \approx n/\beta$, while it goes to zero for small as well as larger $x$.
Introducing the variable $m = n - 1$, Eq.~(\ref{fn3}) can be rewritten as
\begin{eqnarray}
  \label{f5}
  f(x) 
  \!&=&\! 
  \beta \exp(-\beta x) \int dm \, \exp[-\phi(m)] \, ,
  \\
  \label{f6}
  \phi(m) 
  \!&=&\! 
  -\!\ln[P(m \!+\! 1)] - m \ln(\beta x) + \ln[\,\Gamma(m \!+\!1)] \, . \quad
\end{eqnarray}
This integral can be estimated through the saddle-point approximation
expanding $\phi(m)$ to the second order in $\epsilon = m - m_0$, where 
$m_0 = m_0(x)$ locates the maximum of $\phi(m)$, 
defined by $\phi'(m_0) = 0$ and $\phi''(m_0) > 0$,
and integrating over the whole range of $m$,
\begin{eqnarray}
  \label{f5b}
  f(x) 
  &\approx&
  \beta \exp[-\beta x - \phi(m_0)]
  \int_{-\infty}^{+\infty} d\epsilon \, 
  \exp[- \phi''(m_0) \epsilon^2 / 2]
  \nonumber \\
  &=&
  \beta \sqrt{ \frac{2\pi}{\phi''(m_0)} } 
  \exp[-\beta x - \phi(m_0)] \, .
\end{eqnarray}
In order to find $m_0$ we use the Stirling approximation~\cite{Gamma} in 
Eq.~(\ref{f6}), $\Gamma(m+1) \approx \sqrt{\, 2\pi m \,} (m/e)^m$, so that
\begin{eqnarray}
  \label{f6-large-n}
  &&\phi(m)   \approx   - \ln[P(m \!+\! 1)]  
                        - m \ln(\beta x) 
   \nonumber \\
  &&~~~~~~~~~~~       + \ln(\sqrt{2\pi}) 
                      + \left(\! m \!+\! \frac{1}{2} \! \right) \! \ln(m) 
                      - m 
          , \qquad
   \\
  \label{phi2}
  &&\phi'(n) \approx  - \frac{P'(m \!+\! 1)}{P(m \!+\! 1)}
                      - \ln(\beta x)  
                      + \frac{1}{2m} 
                      + \ln(m) \, ,
   \\
  \label{phi3}
  &&\phi''(n) \approx \frac{P'^2(m\!+\!1)}{P^2(m\!+\!1)}
                      - \frac{P''(m\!+\!1)}{P(m\!+\!1)}
                      - \frac{1}{2m^2} 
                      + \frac{1}{m} \, . \quad
\end{eqnarray}
From Eq.~(\ref{phi3}) the condition $\phi''(m) > 0$ for the existence of a 
maximum is fulfilled for large $m$, if one can neglect the terms 
containing $P$ respect to $1/m$; this can be done for general shapes of $P(n)$
which decrease fast enough. 
 
From Eq.~(\ref{phi2}) in the same limit one can neglect $P'/P$ and $1/m$ respect 
to $\ln(m)$ and the approximate solution of $\phi'(m_0)=0$ is 
$m_0(x) \approx \beta x$.
It can be checked that even keeping higher orders in $1/m$ in Eq.~(\ref{phi2}) 
the asymptotic solution reduces to $m_0(x) = \beta x$ for $\beta x \gg 1$. 
Finally, setting $m = m_0(x) = \beta x$ and using Eqs.~(\ref{f6-large-n}) 
and (\ref{phi3}) in Eq.~(\ref{f5b}), one finds
\begin{eqnarray}
  \label{f2}
  f(x \! \gg \! \beta^{-1})
  \equiv
  f_2(x) = \beta P(1 + \beta x) \, .
\end{eqnarray}
This relation provides the asymptotic form of the density $f(x)$ directly
in terms of the dimension density $P(n)$, in the hypothesis that 
$P(n)$ decreases with $n$ at least as $1/n$.

The approximate form of $f(x)$ at $x \!\ll\! \beta^{-1}$
depends on the details of $P(n)$ at small $n$.
For the same form $P(n) = P_\alpha(n)$ considered above
and setting $\phi(n) \!\approx\! \phi(1) + \phi'(1) (n - 1)$
in (\ref{f5}) and (\ref{f6}), one has
\begin{eqnarray}
  \label{f1}
  \!\!\!\!\!\!&& f(x \!\ll\! \beta^{-1}) 
     \!\equiv\! 
           f_1(x) 
     \!=\! \beta \!\! \int_{0}^{+\infty} 
           \!\!\!\!\!\! dm \, \exp[- \phi(0) -\phi'(0) m - \beta x]
  \nonumber \\
  \!\!\!\!\!\!&&=
  \frac{\beta P(1) \exp(-\beta x) }{- \ln(\beta x) - \gamma - P'(1)/P(1)} .
\end{eqnarray}
Here, from Eq.~(\ref{f6}), we set $\phi(0) = \ln[P(1)]$ and
$\phi'(0) = - \, \gamma - \ln(\beta x) -P'(1)/P(1)$, with
$\gamma = \psi(1) \equiv (d\ln[\Gamma(m)]/dm)_{m=1} \approx 0.57721$
the Euler $\gamma$-constant~\cite{Gamma}.

In Fig.~\ref{fig_1} the function $f_2(x)$ (triangles), given by Eq.~(\ref{f2}), 
is compared at large $x$ with the exact distribution $f(x)$ obtained by 
numerical integration of Eq.~(\ref{fn3}) (continuous lines) for the values 
$\alpha = 1, 2$ for the power law density $P_\alpha(n)$. 
Also the corresponding density $f_1(x)$ (circles), given by Eq.~(\ref{f1}), 
is shown at small $\beta x$.

\emph{Conclusion.}
Canonical statistical mechanics predicts that the equilibrium distribution 
$f(x)$ of a conserved quantity $x$ flowing through a system composed 
of subsystems with different (possibly continuous) dimensions $n$,
distributed according to a dimension density $P(n)$,
can exhibit a robust power law tail.
In fact, a general yet simple formula holds at large $x$, 
which directly relates the equilibrium probability density $f(x)$ 
to the dimension density $P(n)$,
\begin{eqnarray}
  \label{fP1}
  f(x) \approx  \beta P(\beta x) \, , ~~~~~~(x \gg \beta^{-1})
\end{eqnarray}
where $\beta$ represents the system temperature.
In particular, it follows that a dimension density $P(n)$ with 
a power law tail will produce a corresponding power law tail 
in the equilibrium distribution $f(x)$.
The mechanism discussed in this paper may represent a prototype model 
of the effect of a high heterogeneous dimensionality
on the equilibrium properties of many complex systems,
such as gene-regulation network models, 
network-based model of language, 
and kinetic wealth exchange models of economy.
We have discussed how such a mechanism is relevant 
for diffusion processes across complex networks;
in the special case of free diffusion on a non-weighted network
the relation (\ref{fP1}) becomes an identity, if $f(x)$ is identified 
with the load distribution, $P(n)$ with the degree distribution,
and the temperature $\beta^{-1}$ with the average flux $\bar{x}$.
Also, the same mechanism is probably in action in systems which can be modeled 
in terms of kinetic exchange models, where a set of basic units exchange 
randomly a conserved quantity $x$.
In the latter case, the inverse of the dimension $1/n$ represents an effective
interaction strength, since the fraction of quantity $x$ exchanged 
in a pair-wise interaction in a kinetic exchange model is of the order of $1/n$,
as in the case of an energy exchange during binary collisions 
in a fluid in $n$ dimensions.

\begin{acknowledgments}
M.P. acknowledges financial support from 
the Estonian Ministry of Education and Research through
Project No. SF0690030s09,
the Estonian Science Foundation 
through Grant No. 7466, and EU NoE BioSim, LSHB-CT-2004-005137.
A.C. is very grateful to A. Jedidi and N. Millot 
for stimulating discussions and useful criticism.
\end{acknowledgments}


\begin{thebibliography}{22}
\expandafter\ifx\csname natexlab\endcsname\relax\def\natexlab#1{#1}\fi
\expandafter\ifx\csname bibnamefont\endcsname\relax
  \def\bibnamefont#1{#1}\fi
\expandafter\ifx\csname bibfnamefont\endcsname\relax
  \def\bibfnamefont#1{#1}\fi
\expandafter\ifx\csname citenamefont\endcsname\relax
  \def\citenamefont#1{#1}\fi
\expandafter\ifx\csname url\endcsname\relax
  \def\url#1{\texttt{#1}}\fi
\expandafter\ifx\csname urlprefix\endcsname\relax\def\urlprefix{URL }\fi
\providecommand{\bibinfo}[2]{#2}
\providecommand{\eprint}[2][]{\url{#2}}

\bibitem[{\citenamefont{Newman}(2005)}]{Newman2005a}
\bibinfo{author}{\bibfnamefont{M.~E.~J.} \bibnamefont{Newman}},
  \bibinfo{journal}{Contemp. Phys.} \textbf{\bibinfo{volume}{46}},
  \bibinfo{pages}{323} (\bibinfo{year}{2005}).

\bibitem[{\citenamefont{Turcotte}(1999)}]{Turcotte1999a}
  \bibinfo{author}{\bibfnamefont{D.~L.} \bibnamefont{Turcotte}},
  \bibinfo{journal}{Rep. Prog. Phys.} 
  \textbf{\bibinfo{volume}{62}},
  \bibinfo{pages}{1377} 
  (\bibinfo{year}{1999}).
  \bibinfo{author}{\bibnamefont{H.~J.~Jensen}},
  \emph{\bibinfo{title}{Self-Organized Criticality: Emergent Complex Behavior
  in Physical and Biological Systems}} 
  (\bibinfo{publisher}{Cambridge University Press}, 
  \bibinfo{address}{Cambridge}, 
  \bibinfo{year}{1998}).

\bibitem[{\citenamefont{Sornette}(1998)}]{Sornette1998a}
  \bibinfo{author}{\bibfnamefont{D.}~\bibnamefont{Sornette}},
  \bibinfo{journal}{Phys. Rev. E} \textbf{\bibinfo{volume}{57}},
  \bibinfo{pages}{4811} (\bibinfo{year}{1998}).

\bibitem[{\citenamefont{Tsallis}(1988)}]{Tsallis1988}
  \bibinfo{author}{\bibfnamefont{C.}~\bibnamefont{Tsallis}}, \bibinfo{journal}{J.
  Stat. Phys.} \textbf{\bibinfo{volume}{52}}, \bibinfo{pages}{479}
  (\bibinfo{year}{1988}).
  \bibinfo{author}{\bibnamefont{E.M.F.Curado}} \bibnamefont{and}
  \bibinfo{author}{\bibnamefont{C.Tsallis}}, \bibinfo{journal}{J. Phys. A}
  \textbf{\bibinfo{volume}{24}}, \bibinfo{pages}{L69} (\bibinfo{year}{1991}).

\bibitem[{\citenamefont{Treumann and Jaroschek}(2008)}]{Treumann2008a}
  \bibinfo{author}{\bibfnamefont{R.~A.} \bibnamefont{Treumann}} \bibnamefont{and}
  \bibinfo{author}{\bibfnamefont{C.~H.} \bibnamefont{Jaroschek}},
  \bibinfo{journal}{Phys. Rev. Lett.} \textbf{\bibinfo{volume}{100}},
  \bibinfo{pages}{155005} (\bibinfo{year}{2008}).

\bibitem[{\citenamefont{Beck}(2006)}]{Beck2006a}
  \bibinfo{author}{\bibfnamefont{C.}~\bibnamefont{Beck}},
  \bibinfo{journal}{Physica A} \textbf{\bibinfo{volume}{365}},
  \bibinfo{pages}{96} (\bibinfo{year}{2006}).

\bibitem[{\citenamefont{Mungan et~al.}(2005)\citenamefont{Mungan, Kabakcioglu,
  Balcan, and Erzan}}]{Mungan2005a}
  \bibinfo{author}{\bibfnamefont{M.}~\bibnamefont{Mungan}},
  \bibinfo{author}{\bibfnamefont{A.}~\bibnamefont{Kabakcioglu}},
  \bibinfo{author}{\bibfnamefont{D.}~\bibnamefont{Balcan}}, 
  \bibnamefont{and}
  \bibinfo{author}{\bibfnamefont{A.}~\bibnamefont{Erzan}}, 
  \bibinfo{journal}{J. Phys. A} 
  \textbf{\bibinfo{volume}{38}}, 
  \bibinfo{pages}{9599}
  (\bibinfo{year}{2005}).
  \bibinfo{author}{\bibfnamefont{Y.}~\bibnamefont{Seng\"un}} 
  \bibnamefont{and}
  \bibinfo{author}{\bibfnamefont{A.}~\bibnamefont{Erzan}},
  \bibinfo{journal}{Physica A} 
  \textbf{\bibinfo{volume}{365}},
  \bibinfo{pages}{446} 
  (\bibinfo{year}{2006}).

\bibitem[{\citenamefont{Kurchan and Laloux}(1996)}]{Kurchan1996a}
  \bibinfo{author}{\bibfnamefont{J.}~\bibnamefont{Kurchan}} \bibnamefont{and}
  \bibinfo{author}{\bibfnamefont{L.}~\bibnamefont{Laloux}},
  \bibinfo{journal}{J. Phys. A: Math. Gen.} \textbf{\bibinfo{volume}{29}},
  \bibinfo{pages}{1929} (\bibinfo{year}{1996}).

\bibitem[{\citenamefont{Noh and Rieger}(2004)}]{Noh2004a}
  \bibinfo{author}{\bibfnamefont{J.}~\bibnamefont{Noh}} \bibnamefont{and}
  \bibinfo{author}{\bibfnamefont{H.}~\bibnamefont{Rieger}},
  \bibinfo{journal}{Phys. Rev. Lett.} \textbf{\bibinfo{volume}{92}},
  \bibinfo{pages}{118701} (\bibinfo{year}{2004}).

\bibitem{note1}
  This dimension locally estimates the connectivity of a link, differently from 
  other global properties of complex networks, such as the spectral dimension.

\bibitem[{\citenamefont{Masucci and Rodgers}(2009)}]{Masucci2009a}
  \bibinfo{author}{\bibfnamefont{A.}~\bibnamefont{Masucci}} \bibnamefont{and}
  \bibinfo{author}{\bibfnamefont{G.}~\bibnamefont{Rodgers}},
  \bibinfo{journal}{Adv. Compl. Syst.} \textbf{\bibinfo{volume}{12}},
  \bibinfo{pages}{113} (\bibinfo{year}{2009}).

\bibitem[{\citenamefont{Chatterjee and Chakrabarti}(2007)}]{Chatterjee2007b}
  \bibinfo{author}{\bibfnamefont{A.}~\bibnamefont{Chatterjee}} 
  \bibnamefont{and}
  \bibinfo{author}{\bibfnamefont{B.}~\bibnamefont{Chakrabarti}},
  \bibinfo{journal}{Eur. Phys. J. B} 
  \textbf{\bibinfo{volume}{60}},
  \bibinfo{pages}{135} 
  (\bibinfo{year}{2007}).
  \bibinfo{author}{\bibfnamefont{M.}~\bibnamefont{Patriarca}},
  \bibinfo{author}{\bibfnamefont{E.}~\bibnamefont{Heinsalu}}, \bibnamefont{and}
  \bibinfo{author}{\bibfnamefont{A.}~\bibnamefont{Chakraborti}},
  \bibinfo{journal}{Eur. J. Phys. B, submitted}  
  (\bibinfo{year}{2009}).
  \bibinfo{editor}{\bibfnamefont{A.}~\bibnamefont{Chatterjee}},
  \bibinfo{editor}{\bibfnamefont{S.}~\bibnamefont{Yarlagadda}},
  \bibnamefont{and} \bibinfo{editor}{\bibfnamefont{B.~K.}\bibnamefont{Chakrabarti}}, 
  eds., 
  \emph{\bibinfo{title}{Econophysics of Wealth Distributions - Econophys-Kolkata I}} 
  (\bibinfo{publisher}{Springer},\bibinfo{year}{2005}).

\bibitem[{\citenamefont{Chakraborti and Chakrabarti}(2000)}]{Chakraborti2000a}
  \bibinfo{author}{\bibfnamefont{A.}~\bibnamefont{Chakraborti}} \bibnamefont{and}
  \bibinfo{author}{\bibfnamefont{B.~K.} \bibnamefont{Chakrabarti}},
  \bibinfo{journal}{Eur. Phys. J. B} \textbf{\bibinfo{volume}{17}},
  \bibinfo{pages}{167} (\bibinfo{year}{2000}).

\bibitem[{\citenamefont{Mandelbrot}(1960)}]{Mandelbrot1960a}
  \bibinfo{author}{\bibfnamefont{B.}~\bibnamefont{Mandelbrot}},
  \bibinfo{journal}{Int. Econ. Rev.} \textbf{\bibinfo{volume}{1}},
  \bibinfo{pages}{79} (\bibinfo{year}{1960}).

\bibitem[{\citenamefont{Patriarca
  et~al.}(2004{\natexlab{a}})\citenamefont{Patriarca, Chakraborti, and
  Kaski}}]{Patriarca2004a}
  \bibinfo{author}{\bibfnamefont{M.}~\bibnamefont{Patriarca}},
  \bibinfo{author}{\bibfnamefont{A.}~\bibnamefont{Chakraborti}},
  \bibnamefont{and} \bibinfo{author}{\bibfnamefont{K.}~\bibnamefont{Kaski}},
  \bibinfo{journal}{Phys. Rev. E} \textbf{\bibinfo{volume}{70}},
  \bibinfo{pages}{016104} (\bibinfo{year}{2004}{\natexlab{a}}).

\bibitem[{\citenamefont{Chakraborti and Patriarca}(2008)}]{Chakraborti2008a}
  \bibinfo{author}{\bibfnamefont{A.}~\bibnamefont{Chakraborti}} \bibnamefont{and}
  \bibinfo{author}{\bibfnamefont{M.}~\bibnamefont{Patriarca}},
  \bibinfo{journal}{Pramana J. Phys} \textbf{\bibinfo{volume}{71}},
  \bibinfo{pages}{233} (\bibinfo{year}{2008}).

\bibitem[{\citenamefont{Patriarca
  et~al.}(2004{\natexlab{b}})\citenamefont{Patriarca, Chakraborti, and
  Kaski}}]{Patriarca2004b}
  \bibinfo{author}{\bibfnamefont{M.}~\bibnamefont{Patriarca}},
  \bibinfo{author}{\bibfnamefont{A.}~\bibnamefont{Chakraborti}},
  \bibnamefont{and} \bibinfo{author}{\bibfnamefont{K.}~\bibnamefont{Kaski}},
  \bibinfo{journal}{Physica A} \textbf{\bibinfo{volume}{340}},
  \bibinfo{pages}{334} (\bibinfo{year}{2004}{\natexlab{b}}).

\bibitem[{\citenamefont{Abramowitz and Stegun}(1970)}]{Gamma}
  \bibinfo{editor}{\bibfnamefont{M.}~\bibnamefont{Abramowitz}} \bibnamefont{and}
  \bibinfo{editor}{\bibfnamefont{I.~A.} \bibnamefont{Stegun}}, eds.,
  \emph{\bibinfo{title}{Handbook of Mathematical Functions}}
  (\bibinfo{publisher}{Dover}, \bibinfo{address}{N.Y.}, \bibinfo{year}{1970}).

\end{thebibliography}
\end{document}